\documentclass{article}
\pdfoutput=1 

\usepackage{arxiv}

\usepackage[utf8]{inputenc} 
\usepackage[T1]{fontenc}    
\usepackage{hyperref}       
\usepackage{url}            
\usepackage{booktabs}       
\usepackage{amsfonts}       
\usepackage{nicefrac}       
\usepackage{microtype}      
\usepackage{lipsum}		
\usepackage{graphicx}
\usepackage{doi}

\usepackage{amsmath, amsfonts, bbm, bm}
\usepackage{authblk}
\usepackage{graphicx}
\usepackage{chemmacros}
\usepackage{algorithm,algcompatible}
\algnewcommand\algorithmicreturn{\textbf{return}}
\algnewcommand\RETURN{\algorithmicreturn}
\algnewcommand\algorithmicprocedure{\textbf{procedure}}
\algnewcommand\PROCEDURE{\item[\algorithmicprocedure]}%
\algnewcommand\algorithmicendprocedure{\textbf{end procedure}}
\algnewcommand\ENDPROCEDURE{\item[\algorithmicendprocedure]}%
\algnewcommand{\algvar}[1]{{\text{\ttfamily\detokenize{#1}}}}
\algnewcommand{\algarg}[1]{{\text{\ttfamily\itshape\detokenize{#1}}}}
\algnewcommand{\algproc}[1]{{\text{\ttfamily\detokenize{#1}}}}
\algnewcommand{\algassign}{\leftarrow}
\DeclareMathOperator*{\argmax}{arg\,max}

\title{Evaluating the accuracy of Gaussian approximations in VSWIR imaging spectroscopy retrievals}

\author[1]{Kelvin M. Leung}
\author[2]{David R. Thompson}
\author[2]{Jouni Susiluoto}
\author[1]{Jayanth Jagalur-Mohan}
\author[2]{Amy Braverman}
\author[1]{Youssef Marzouk}
\affil[1]{Massachusetts Institute of Technology}
\affil[2]{NASA Jet Propulsion Laboratory}

\date{}

\renewcommand{\headeright}{K. Leung, D. Thompson, J. Susiluoto, J. Jagalur-Mohan, A. Braverman, Y. Marzouk}
\renewcommand{\undertitle}{}
\renewcommand{\shorttitle}{}

\hypersetup{
pdftitle={A template for the arxiv style},
pdfsubject={q-bio.NC, q-bio.QM},
pdfauthor={David S.~Hippocampus, Elias D.~Striatum},
pdfkeywords={First keyword, Second keyword, More},
}

\begin{document}
\maketitle

\begin{abstract}

The joint retrieval of surface reflectances and atmospheric parameters in VSWIR imaging spectroscopy is a computationally challenging high-dimensional problem. 
Using NASA's Surface Biology and Geology mission as the motivational context, the uncertainty associated with the retrievals is crucial for further application of the retrieved results for environmental applications.
Although Markov chain Monte Carlo (MCMC) is a Bayesian method ideal for uncertainty quantification, the full-dimensional implementation of MCMC for the retrieval is computationally intractable.

In this work, we developed a block Metropolis MCMC algorithm for the high-dimensional VSWIR surface reflectance retrieval that leverages the structure of the forward radiative transfer model to enable tractable fully Bayesian computation.
We use the posterior distribution from this MCMC algorithm to assess the limitations of optimal estimation, the state-of-the-art Bayesian algorithm in operational retrievals \cite{oe} which is more computationally efficient but uses a Gaussian approximation to characterize the posterior.
Analyzing the differences in the posterior computed by each method, the MCMC algorithm was shown to give more physically sensible results and reveals the non-Gaussian structure of the posterior, specifically in the atmospheric aerosol optical depth parameter and the low-wavelength surface reflectances.
\end{abstract}


\section{Introduction}

Whether airborne or orbital, all remote sensing missions face a common challenge of characterizing distant objects using only measurements made at the sensor.
In the Earth sciences, investigators will often solve this problem with physics-based models that use the state of the surface or atmosphere to predict the remote measurement.
Investigators can then {\it retrieve}, or determine, the state most consistent with the remote data.  
Model inversion methods are used for diverse sensors ranging from infrared or microwave sounders, to multiangle imagers, to radiometers. 
Perhaps one of the most challenging applications from a computational perspective is remote Visible/Short-Wave Infrared (VSWIR) imaging spectroscopy \cite{esas2018}.  

VSWIR imaging spectrometers acquire a data cube with two spatial dimensions and one spectral dimension.  
In other words, they produce images in which each pixel contains a radiance spectrum covering the entire solar reflected interval from 380 to 2500 nm.  
This interval is sensitive to diverse surface and atmospheric processes, making these sensors useful for a wide range of studies from terrestrial and aquatic ecology, to geology, to hydrology and cryosphere studies.  
These Earth surface studies aim to measure properties of the surface that create characteristic features in the reflectance spectra. 
Roughly speaking, surface reflectance is the fraction of incident illumination at the surface which is reflected back in the direction of the sensor.  
However, imaging spectrometers observe radiance at the top of the atmosphere, so inference to remove the effects of the atmosphere is required to estimate surface reflectance at each pixel.
The reflectance can then be used to further estimate properties of the Earth surface. 
Because of the high data volume of these sensors and their broad spectral range encompassing a wide range of physical phenomena, VSWIR imaging spectrometers present a particularly challenging test case for efficient inference algorithms.

Our motivating context for this problem is NASA's Surface Biology and Geology mission (SBG) \cite{sbgreview, sbg2}.
The objective of SBG is to track changes in surface properties pertaining to ecosystems, coastal zones, agriculture, and snow and ice accumulations over time, for the entire planet, by first retrieving the surface reflectances.
For these types of scientific applications, the uncertainty associated with the retrieval is particularly important.
This motivates the need for a Bayesian method to determine the posterior distribution of the surface reflectances and related atmospheric parameters.

Markov chain Monte Carlo (MCMC) is a Bayesian sampling method that was introduced in the context of remote sensing retrievals in \cite{tamminen2001, tamminen2004, haario2004}.
Recent retrieval problems are generally high-dimensional and require methods of dimension reduction to lower the computational complexity. 
For example, \cite{wang2013} breaks up the high-dimensional parameter space into low-dimensional blocks that can be sampled in parallel.
\cite{haario2004} and \cite{tukiainen2016} implement MCMC in a low-dimensional parameter subspace obtained from principal component analysis (PCA). 

More recent methods of dimension reduction such as likelihood informed subspace (LIS) have also been considered in the retrieval context, where MCMC is performed in a specific low-dimensional subspace that is determined by the data \cite{LIS}.
\cite{ottoCH4} uses LIS in atmospheric methane retrievals, and \cite{ottoOCO2} uses LIS in retrievals for atmospheric concentrations of carbon dioxide in NASA's Orbiting Carbon Observatory-2 (OCO-2) mission.
Contrary to most retrievals where the number of parameters is under 100, there are over 400 parameters in the SBG retrieval problem, which makes the problem much more difficult in terms of computational tractability. 
Furthermore, since dimension reduction methods such as PCA or LIS are low-dimensional approximations, significantly reducing the dimension leads to problems in convergence to the posterior distribution.
In this work, we present a method of utilizing the structure of the problem to overcome the high dimensionality and create a tractable sampling algorithm.

Optimal estimation (OE) \cite{oe} is the current state-of-the-art algorithm for an operational setting, such as for NASA's EMIT mission \cite{emit1, emit2}.
OE is a Bayesian retrieval algorithm that computes a maximum a posteriori (MAP) estimate of the parameters and characterizes the posterior distribution using the Laplace approximation.
Although this leads to fast Gaussian approximations of the posterior, the posterior is not Gaussian in the general case.

Our objective is to explore how well this approximation holds.
There are two main contributions of this work.
\begin{enumerate}
    \item We developed a computationally tractable Block Metropolis MCMC algorithm for the VSWIR retrieval problem. This fully Bayesian algorithm allows for the characterization of a non-Gaussian posterior and performs exact inference in the limit of infinite samples.
    \item We use this algorithm to evaluate the limitations of the OE method and identify the scenarios in which it is sufficiently accurate.
\end{enumerate}

\section{Setup of the remote sensing problem}
\label{setup}

The remote sensing retrieval considered in this paper is modelled as an inverse problem. 
For each pixel of the image captured by the imaging spectrometer, the quantities of interest are the surface and atmospheric parameters that are retrieved given the radiance at the same pixel.
This type of retrieval can be thought of as a statistical inference problem for one set of multidimensional parameters with one set of multidimensional data.

We use the notation $\mathbf{y}$ to denote the set of radiance observations from the imaging spectrometer.
The radiances are used to infer the state $\mathbf{x}$, which consists of the surface and atmospheric parameters.
Incoming solar radiation is reflected off the Earth surface, and the transfer of radiation through the atmosphere is modelled by a vector-valued forward function $f(\cdot)$.
The full expression is known and is written in \eqref{eq:fwd}.
The observations are represented by the output of the forward model with additive noise, $\mathbf{y} = f(\mathbf{x}) + \epsilon$.

The setup of this inference problem and the desire for uncertainty quantification leads to a Bayesian formulation.
Given the prior and likelihood distributions, the posterior distribution of the surface and atmospheric parameters conditioned on the observed radiance, $\pi(\mathbf{x}|\mathbf{y})$, is obtained using Bayes rule:
\begin{equation}
    \pi(\mathbf{x}|\mathbf{y}) = \frac{\pi(\mathbf{y}|\mathbf{x}) \pi(\mathbf{x})}{\pi(\mathbf{y})} \propto \pi(\mathbf{y}|\mathbf{x})\pi(\mathbf{x}).
\end{equation}
This section provides an overview of the elements associated with the prior and likelihood distributions, including parameters and data.
Formulations to determine the posterior distribution are described in Section~\ref{formulations}.

\subsection{Surface and atmospheric parameters}

The inversion problem estimates the surface and atmospheric parameters, which are concatenated into one state vector: 
\begin{equation}
    \mathbf{x} = [\mathbf{x}_{\text{refl}}, \mathbf{x}_{\text{atm}}]^\top,
\end{equation}
where $\mathbf{x}_{\text{refl}} \in \mathbbm{R}^n$ and  $\mathbf{x}_{\text{atm}} \in \mathbbm{R}^2$.
There are $n=432$ surface parameters and two atmospheric parameters.
\begin{itemize}
    \item The surface parameters $\mathbf{x}_{\text{refl}}$ are surface reflectances that describe the proportion of solar radiation that is reflected off the surface at each of the $n$ wavelengths.
The wavelengths range from 380 to 2500 nm. 
    \item The two atmospheric parameters are $\mathbf{x}_\text{atm} = [x_\text{AOD}, x_\text{H2O}]^\top$, which consist of Aerosol Optical Depth (AOD) at 550 nm and column precipitable water vapour (cm).
\end{itemize}

Aerosol optical depth (AOD) is a measure of the atmospheric concentration of aerosols. Specifically, it is the proportion of radiation that is absorbed by aerosols at wavelength 550 nm.
The second atmospheric parameter is the column precipitable water vapour (cm), which is the volume of water per vertical column of atmosphere. 

\textbf{Prior.}
The prior on the parameters is modelled as a normal distribution.
The surface and atmospheric parameters are treated independently, giving a block diagonal structure to the prior covariance:

\begin{equation}
    \bm{\mu}_\text{pr} = \big[\bm{\mu}^0_{\text{refl}}, \bm{\mu}^0_{\text{atm}} \big]^\top,\,\,\,\, \Gamma_\text{pr} = 
    \begin{bmatrix}
        \Gamma^0_{\text{refl}} & 0\\
        0 & \Gamma^0_{\text{atm}}
    \end{bmatrix}
\end{equation}

The surface prior is created using a set of over 1400 historical reflectance spectra from the EcoSIS spectral library \cite{team2020ecological}.
These spectra are fitted to a Gaussian mixture model with eight components, each corresponding to a different type of terrain on the Earth surface that have similar characteristics, such as vegetation or aquatic environments. 
For a particular inversion, the component with shortest Mahalanobis distance to the initial state estimate is used as the Gaussian surface prior.

For the atmospheric parameters, the priors are chosen to have large variances to allow relatively unconstrained exploration of the posterior.

\subsection{Forward model}

The forward model, $f(\cdot) = [f_1(\cdot), \dots , f_n(\cdot)]^\top$, approximates the propagation of photons through the Earth atmosphere from the surface to the imaging spectrometer.
The model used in this work consists of two stages to map the state $\mathbf{x}$ to the radiance $\mathbf{y}$.

The first step is the computation of intermediate parameters using MODTRAN 6.0, a high-fidelity radiative transfer model \cite{modtran}.
The outputs of MODTRAN given $\mathbf{x}_\text{atm}$ are three $n$-dimensional parameters that describe light propagation through the atmosphere: path reflectance $\bm{\rho}_a$, spherical albedo $\mathbf{s}$, and atmospheric transmission $\mathbf{t}$. 
For computational efficiency, a lookup table is generated for a set of reference atmospheric conditions.

The second step is to use the intermediate parameters to calculate the radiance. After linearly interpolating these parameters in the lookup table, the mechanics of the forward model for $i = 1, \dots, n$ are given by:
\begin{equation} \label{eq:fwd}
    f_i(\mathbf{x}) = \frac{\phi_0}{\pi} e_{0,i} \bigg[ \rho_{a,i} (\mathbf{x}_\text{atm}) + \frac{t_i(\mathbf{x}_\text{atm}) \cdot x_{\text{refl},i}}{1 - s_i(\mathbf{x}_\text{atm}) \cdot x_{\text{refl},i}} \bigg], 
\end{equation}
where $\phi_0$ is the cosine of the solar zenith angle and $\mathbf{e_0}$ is the solar downward irradiance at the top of the atmosphere \cite{fwdModelEqn}.




\subsection{Noise model}
We model the observation uncertainty matrix by combining covariance matrices from different independent error sources as in \cite{thompson2020}. First, we determine the intrinsic sensor noise, which includes uncertainty due to discrete photon counts, electronic uncertainty in the analog-to-digital conversion, and thermal noise from the instrument itself. These sources are all independent and uncorrelated in each channel leading to a diagonal covariance structure. The photon noise contribution depends on the magnitude of the radiance itself, so we use the measurement itself to predict its own noise level. Any error induced by this circularity is acceptable since noise is small relative to the total magnitude, with a signal to noise ratio of 500 or 1000 for typical spectra.

In addition to the instrument noise, we also model several systematic error sources following \cite{oe}. We use a diagonal matrix to represent a 1\% uncertainty in calibration. Another diagonal matrix represents systematic radiative transfer model errors due to spectral calibration uncertainty and the intrinsic uncertainty in the unretrieved components of the atmospheric model. The covariance matrices from these independent error sources combine additively to form a final observation error matrix $\Gamma_\text{obs}$.



\subsection{Retrieval}
\begin{figure}
    \centering
    \includegraphics[width=\linewidth]{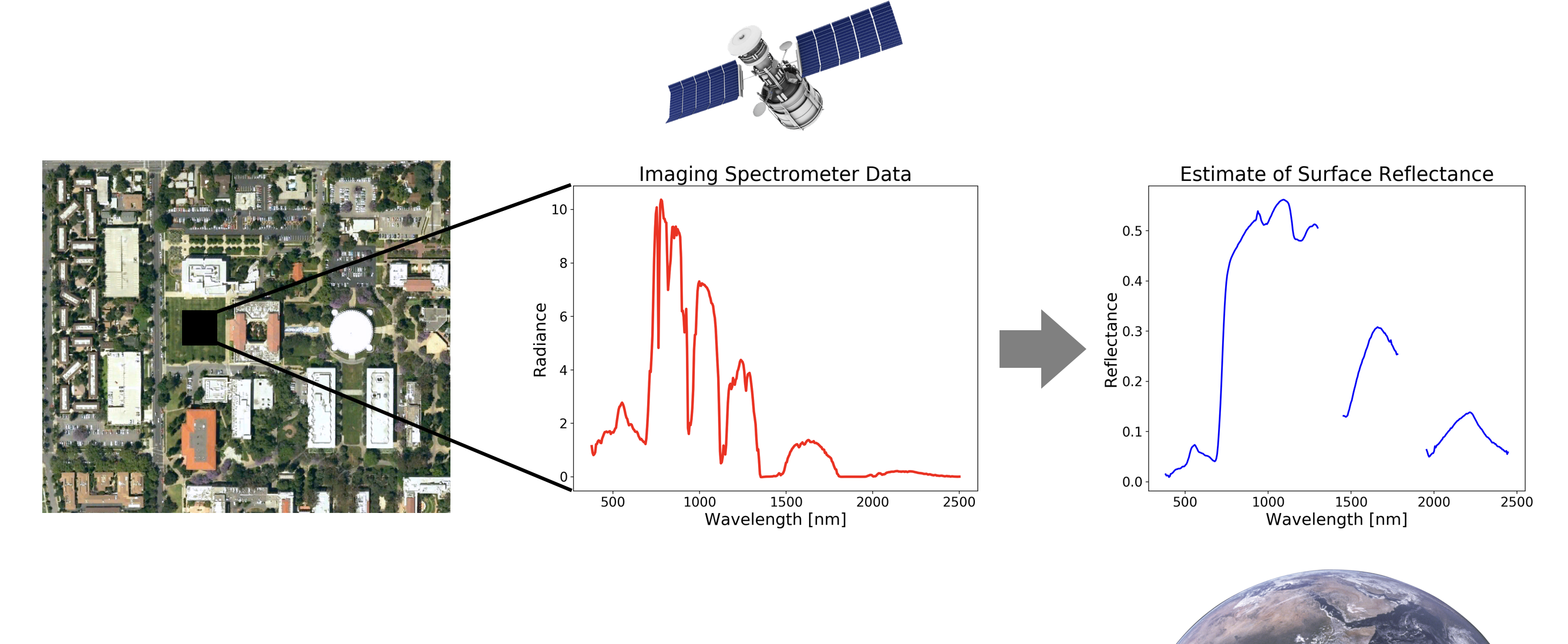}
    \caption{Sample retrieval over a grassy field from radiance to reflectance.} 
    \label{fig:retrieval}
\end{figure}

The prior, forward, and noise models are used to solve the inversion problem of retrieving the surface and atmospheric parameters given the observations. 
From a Bayesian perspective, the retrieval is equivalent to characterizing the posterior that is given by:
\begin{align}
    \pi (\mathbf{x} | \mathbf{y}) \propto \exp \bigg(-\frac{1}{2} \big\| \mathbf{x} - \bm{\mu}_\text{pr} \big\|_{\Gamma_\text{pr}}^2 -   \frac{1}{2} \big\| \mathbf{y} - f(\mathbf{x})  \big\|_{\Gamma_\text{obs}}^2  \bigg) 
\end{align}
Methods for inference are described in Section~\ref{formulations}.
An example of the retrieval process from a pixel in the satellite image to the surface reflectance is shown in Figure~\ref{fig:retrieval}.
The gaps in the reflectance represent the wavelengths for which most of the radiation is absorbed by water droplets in the atmosphere, rendering the retrieval meaningless in those regions. 
Ignoring these wavelengths, there are $324$ reflectance parameters of interest.

\section{Estimation and inference formulations} \label{formulations}


The existing state-of-the-art Bayesian method used in VSWIR remote sensing problems is optimal estimation (OE) \cite{oe}, which is a computationally efficient way to obtain estimates of the surface and atmospheric parameters.
In this work we place the emphasis on the uncertainty quantification of the retrieval of these parameters.
We consider a sampling-based Markov chain Monte Carlo (MCMC) approach to characterize the full posterior distribution.

\subsection{Optimal estimation}
Given a Gaussian prior and the forward and noise models, the OE method solves an optimization problem to estimate the surface and atmospheric parameters. 
The negative log posterior is used as the objective function, and the resulting parameter estimate is denoted as $\mathbf{x}_\text{MAP} = \argmax_{\mathbf{x}}{c(\mathbf{x})}$, where:
\begin{align}
\begin{split}
    c(\mathbf{x}) = \frac{1}{2}(\mathbf{x}-\bm{\mu}_\text{pr})^\top \Gamma_\text{pr}^{-1} (\mathbf{x}-\bm{\mu}_\text{pr}) \\
    + \frac{1}{2}(\mathbf{y} - f(\mathbf{x}))^\top \Gamma_\text{obs}^{-1} (\mathbf{y} - f(\mathbf{x})).
\end{split}
\end{align}

The covariance estimate is a Laplace approximation \cite{laplace1} derived from linear Bayesian inversion theory \cite{optlowrank} using the local linearization of the forward model at the MAP estimate \cite{laplaceForPos}. 
\begin{equation}
    \Gamma_\text{L} = \bigg( \nabla f(\mathbf{x}_\text{MAP})^\top \Gamma_\text{obs}^{-1} \nabla f(\mathbf{x}_\text{MAP}) + \Gamma_\text{pr}^{-1} \bigg)^{-1}
\end{equation}
Since the OE posterior is characterized using a local Gaussian approximation, we sometimes refer to OE as the \textit{approximate Bayesian} method to contrast with the \textit{fully Bayesian} MCMC approach.

\subsection{Fully Bayesian approach}
Although the Laplace approximation in optimal estimation would be accurate if the posterior is approximately Gaussian, this is not the case in general.
Since the forward model is nonlinear, the posterior shape cannot be determined a priori, making it impossible to determine whether a normal approximation is sufficiently accurate.
Therefore, a method of characterizing the full posterior distribution is needed to obtain an accurate measure of uncertainty associated with the retrieval.

Markov chain Monte Carlo (MCMC) \cite{mcmc} is a probabilistic sampling method that addresses the issue of characterizing the posterior, but is computationally intractable in the high-dimensional VSWIR retrieval problem.
Methods for dimension reduction such as \cite{LIS, certifiedDimRed, activesubspace} designed for Bayesian inverse problems can limit the sampling to a low-dimensional subspace.
However, these methods are considered \textit{approximate} inference because they involve truncation of information that are deemed less important based on the eigenvalues.
This paper presents a technique for \textit{exact} inference specifically for the VSWIR retrieval that takes advantage of the conditional dependence structure of the surface and atmospheric parameters in the state vector.
The next subsection outlines the structure that we then use to develop the sampling methodology in Section~\ref{methodology}.

\subsection{Linear approximations to the forward model} \label{linmod}


The structure in the forward model arises from $\mathbf{x}_\text{refl}$ and $\mathbf{x}_\text{atm}$ being treated as independent in \eqref{eq:fwd}. 
When the atmospheric parameters are held fixed, the forward model is approximately linear. 
That is, for $\mathbf{x}_\text{atm}$ held constant, we can define a submodel conditioned on the atmospheric parameters:
\begin{equation} \label{eq:linmod}
    f_\text{refl} (\mathbf{x}_\text{refl}) \approx \mathbf{A} \mathbf{x}_\text{refl} + \mathbf{b} ,
\end{equation}
where $\mathbf{A}$, a diagonal matrix with $A_{ii} = \frac{\phi_0}{\pi} e_{0,i} t_i$, and $b_i = \frac{\phi_0}{\pi} e_{0,i} \rho_{a,i}$, $i = 1\dots n$, are the deterministic constants computed from the intermediate parameters computed using MODTRAN.
The denominator of the second term in \eqref{eq:fwd} is approximately equal to one.
The approximately linear structure in the surface parameters can be exploited to accelerate the sampling process.

\section{Sampling methodology} \label{methodology}

A computationally tractable fully Bayesian algorithm was developed to obtain samples from the posterior distribution of the surface and atmospheric parameters.
The algorithm, based on a block Metropolis MCMC algorithm \cite{mcmc, mcmcblock}, generates alternating samples of the reflectance and atmospheric parameter blocks.
Contrary to algorithms involving dimension reduction, this algorithm performs exact inference, meaning that it converges to the true posterior distribution in the limit of infinite samples.
The algorithm is described in this section, including the overall structure and the parameter tuning process.

\subsection{Exploiting structure in the forward model}
The forward model is known to be approximately linear in the reflectances conditioned on fixed atmospheric parameters.
Since the objective is to develop a fully Bayesian algorithm, the linear model described in Section~\ref{linmod} is not used explicitly in the inversion.
It is instead used to provide structure to the sampling algorithm.

The motivation behind the block Metropolis algorithm is to restrict the ``difficult" sampling to the atmospheric parameters. After obtaining a sample from the atmospheric block, the sampling within the surface block can converge much faster thanks to the approximate conditional linearity given a fixed atmosphere.
Without this structure, the algorithm would have to blindly explore the $n$-parameter space, which is computationally infeasible in practice.

The sampling procedure is as follows.
The chain is first initialized at the MAP estimate obtained using optimal estimation.
Each subsequent sample is split into the atmospheric and reflectance blocks, each with a proposal and acceptance step.
The proposal is a sample from the normal distribution centered at the previous sample with some proposal covariance.
This proposal covariance is discussed in the next subsection.
The proposal is then accepted or rejected with some acceptance probability computed based on the posterior density.
The new samples from both blocks are then concatenated and added to the chain.
The acceptance step ensures asymptotic convergence of the chain.
The full algorithm is outlined in Section~\ref{sec:alg}.

\subsection{Choice of proposal covariance}

Different approaches were taken for the two blocks since the structure is known for the reflectances conditioned on the atmospheric parameters but not vice versa. 
For the atmospheric block, the proposal covariance follows the update procedure of the Adaptive Metropolis algorithm \cite{mcmcAM}, in which the proposal attempts to adapt to the shape of the posterior based on the previous samples to explore the parameter space more efficiently.
This adaptive scheme is given as:
\begin{equation}
    \Gamma_\text{atm}^{\,(i)}=
    \begin{cases}
        \epsilon_0 \,I_2 & i \leq 1000 \\
        s_2 \, \text{cov} \big(\mathbf{x}_\text{atm}^{(0)}\,, \dots ,\mathbf{x}_\text{atm}^{(i-1)} \big) + s_2\,\epsilon_\text{AM}\, I_2 & i > 1000
    \end{cases},
\end{equation}
where $s_2 = \frac{2.38^2}{2}$ and $\epsilon_0 = \epsilon_\text{AM} = 10^{-3}$.

Two methods of obtaining the proposal covariance for the reflectance block were compared.
Both involve computing some approximation of the posterior covariance of the reflectance parameters.

1) \textbf{Linear inversion theory.}
Modelling the forward submodel as linear, i.e. making \eqref{eq:linmod} an equality, closed form expressions of the posterior covariance can be derived from linear Bayesian inversion theory \cite{optlowrank}.
For the linear model in \eqref{eq:linmod} and using the same prior and noise model, the posterior covariance of the reflectances can be expressed as:
\begin{equation} \label{eq:linpos}
    \Gamma_\text{lin}^{\,(i)} = \big(\mathbf{I} - \Gamma_\text{pr} \,\mathbf{A}^{(i)T}\,\Gamma_y^{\,-1\,(i)} \,\mathbf{A}^{(i)} \big) \,\Gamma_\text{pr},
\end{equation}
where $\Gamma_y^{\,(i)} = \mathbf{A}^{(i)}\, \Gamma_\text{pr} \,\mathbf{A}^{(i)T} + \Gamma_\text{obs}$ is the marginal covariance of the data.
A scaled version of this posterior approximation, $\epsilon_1 \Gamma_\text{lin}$, where $\epsilon_1 < 1$, is used as the proposal covariance.

2) \textbf{Laplace approximation.}
Another method is to directly use the Laplace approximation, $\Gamma_\text{L}$, obtained from OE  \cite{laplace1, laplaceForPos}. 
This can be done as a preprocessing step to avoid having to compute an inversion problem for every iteration of the chain. 
The proposal covariance is the scaled Laplace approximation, $\epsilon_2 \Gamma_\text{L}$, for some $\epsilon_2<1$.

The block Metropolis algorithm was implemented to compare the quality of mixing in the chain for each method of obtaining the proposal covariance.
Two million samples were generated for each chain, which were then thinned by taking every 10 samples for a total of $2\times 10^{5}$ samples.
The mixing characteristics were compared by analyzing trace plots and the effective sample sizes. 
Both scaling parameters $\epsilon_1$ and $\epsilon_2$ were tuned to achieve a near-optimal acceptance rate of approximately 23\% \cite{optscaling0234}.
The subsequent results use the scaling parameters $\epsilon_1=0.14$ and $\epsilon_2=0.11$.

\begin{figure}
    \centering
    \includegraphics[width=0.5\linewidth]{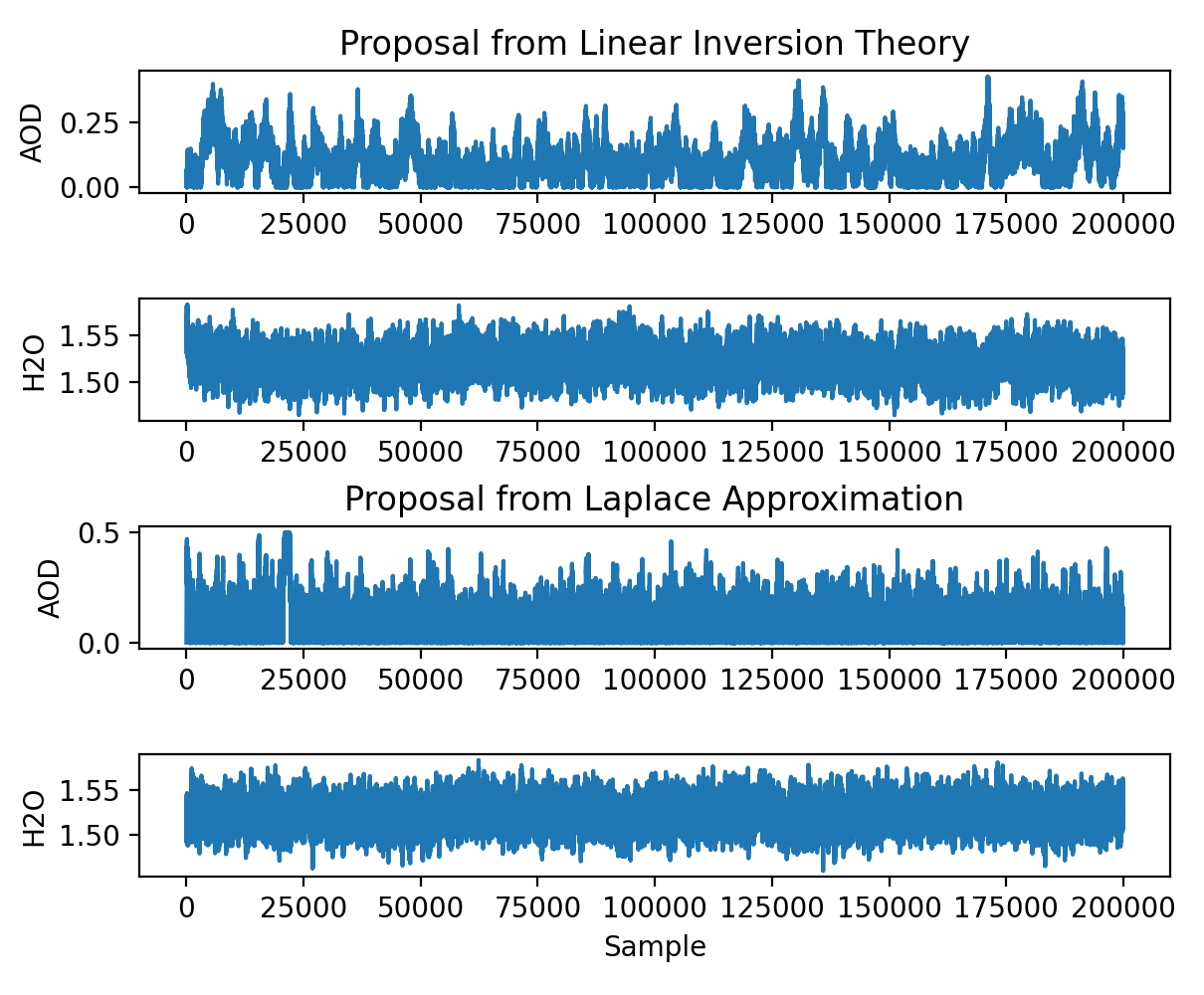}
    \caption{Trace plots of the atmospheric parameters using the two methods of obtaining proposal covariance for the reflectance block.} 
    \label{fig:trace}
\end{figure}

The samples from the reflectance block affect the acceptance in the atmospheric block due to the alternating nature of the sampling process, so we must evaluate the effect of this proposal covariance on the parameters in both blocks.
The trace plots for the atmospheric parameters are shown in Figure~\ref{fig:trace}.
While the \ch{H2O} parameter trace plots are similar across the two methods, the AOD plots show that the chain explores the parameter space more efficiently when using the Laplace approximation. 
In the proposal from linear inversion theory, the chain requires more samples to move toward a different region of the space.

Since MCMC leads to dependent samples, another useful metric is the autocorrelation time $\tau_a$, which is the time it takes for the samples to become effectively independent. 
The effective sample size (ESS) can then be defined as:
\begin{equation}
    \text{ESS} = \frac{N}{\tau_a},
\end{equation}
where $N$ was taken as $1.8 \times 10^5$ after removing the first $2\times 10^4$ samples in the chain as burn-in.

\begin{figure}
    \centering
    \includegraphics[width=0.5\linewidth]{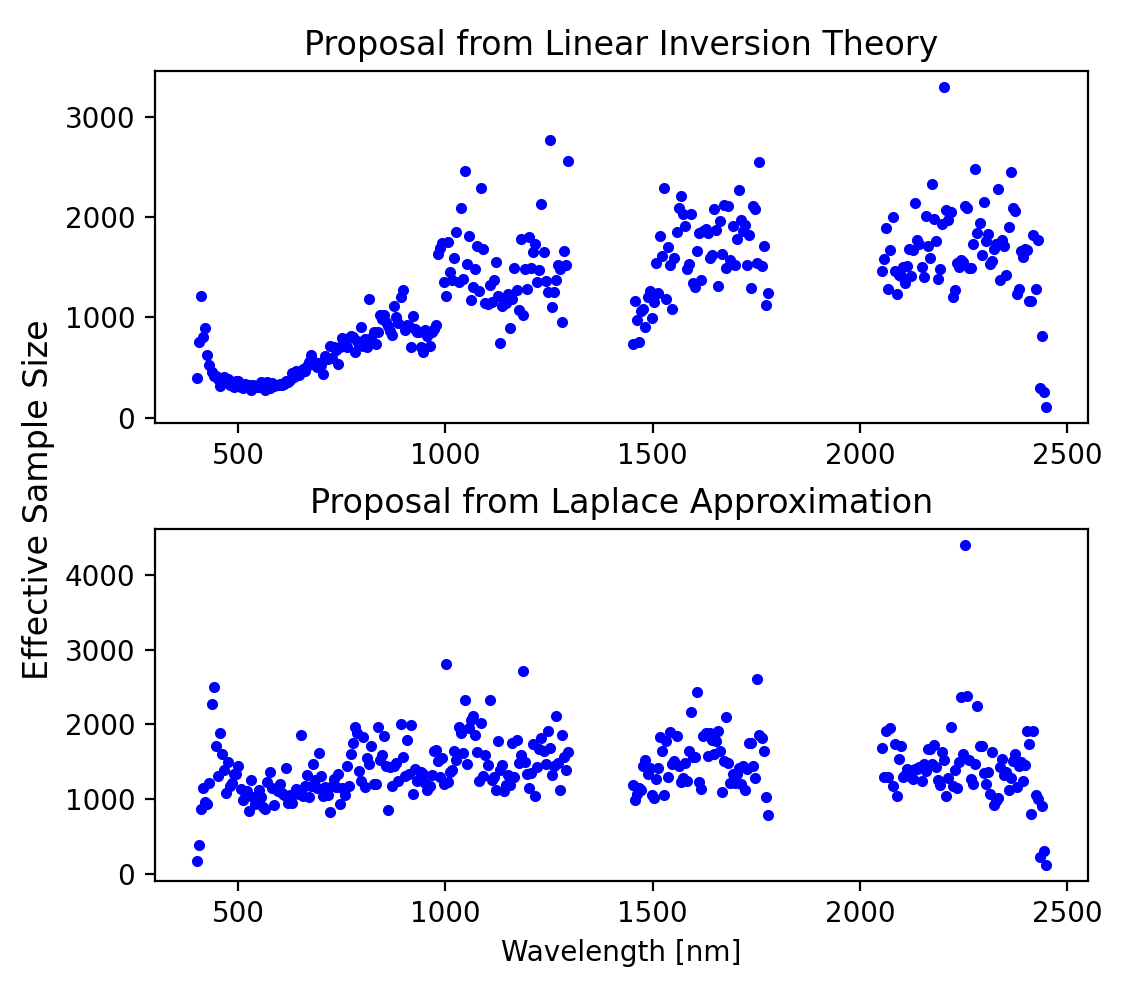}
    \caption{Effective sample sizes of the reflectance chains. } 
    \label{fig:ess}
\end{figure}

The ESS was computed for each of the $n+2$ parameters and summarized in Table~\ref{tab:ess}.
The median value is around 1000, indicating that one effectively independent sample is generated every 180 samples.
The proposal covariance obtained from the Laplace approximation generates greater sample sizes throughout, suggesting better mixing and greater sampling efficiency.
This method performs particularly better for the atmospheric AOD parameter, where the ESS is roughly a factor of 4 greater than the linear inversion method.
For the reflectances, it induces a much better performance for wavelengths less than 1000 nm, as shown in Figure~\ref{fig:ess}.
While the ESS from the Laplace method remain fairly consistent throughout all wavelengths, the ESS in low wavelength regions drop significantly for the linear inversion method.

Another advantage of the Laplace method is that the Laplace approximation is constant and only needs to be computed once as a preprocessing step. 
The linear inversion method requires the computation of \eqref{eq:linpos} between the two blocks in the algorithm for each sample.
Although this does not make a noticeable difference in computational time, it simplifies the algorithm.
In the final implementation of the algorithm, the proposal covariance for the reflectance block is equal to $\epsilon_2 \Gamma_\text{L}$, where the scaling parameter was tuned to be $\epsilon_2 = 0.11$.

\begin{table}
    \centering
    \caption{Effective sample sizes for MCMC on Building 177}
    \label{tab:ess}
    \begin{tabular}{|c|c|c|}
        \hline
        & Proposal from linear inversion & Proposal from Laplace \\
         \hline
        Refl Min & 108 & 120 \\
        \hline
        Refl Med & 1278 & 1375\\
        \hline
        Refl Max & 3294 & 4399 \\
        \hline
        AOD & 166 & 633 \\
        \hline
        \ch{H2O} & 527 & 784 \\ 
         \hline
    \end{tabular}
\end{table}

\subsection{Block Metropolis Algorithm} \label{sec:alg}

Algorithm~\ref{alg:1} outlines the final algorithm, which takes advantage of the forward model structure in which $f_\text{refl}(\cdot)$ is approximately linear.
By using this property along with the scaled Laplace approximation as the proposal covariance, we are able to obtain samples that efficiently explore the parameter space of both the surface and atmospheric parameters.
Note that in the atmospheric block, $\mathbf{z}_\text{atm}$ is drawn from a truncated normal distribution with a lower bound of zero. 

\begin{algorithm}
\caption{Block Metropolis}\label{alg:1}
\begin{algorithmic}[1]
\State Initialize $\mathbf{x}^{(0)} = \mathbf{x}_\text{MAP}$
\FOR{$i = 1\dots N$}
    \State Sample $\mathbf{x}^{(i)}_\text{atm}$
    \State \ \ \ \ Proposal $\mathbf{z}_\text{atm} \sim \mathcal{N}\big(\mathbf{x}^{(i-1)}_\text{atm}, \,\,\Gamma_\text{atm}^{(i)}\big)$ such that $z_\text{atm} \geq 0$
    \State \ \ \ \ Metropolis accept/reject for $\big[\mathbf{x}^{(i-1)}_\text{refl}, \mathbf{z}_\text{atm}\big]$
    \State Sample $\mathbf{x}^{(i)}_\text{refl}$
    \State \ \ \ \ Proposal $\mathbf{z}_\text{refl} \sim \mathcal{N}\big(\mathbf{x}^{(i-1)}_\text{refl}, \,\,\epsilon_2 \,\Gamma_\text{L}\big)$
    \State \ \ \ \ Metropolis accept/reject for $\big[\mathbf{z}_\text{refl}, \mathbf{x}^{(i)}_\text{atm} \big]$
    \State Compute $\Gamma_\text{atm}^{(i+1)}$
\ENDFOR
\end{algorithmic}
\end{algorithm}

\section{Results}

\begin{figure}
    \centering
    \includegraphics[width=0.7\linewidth]{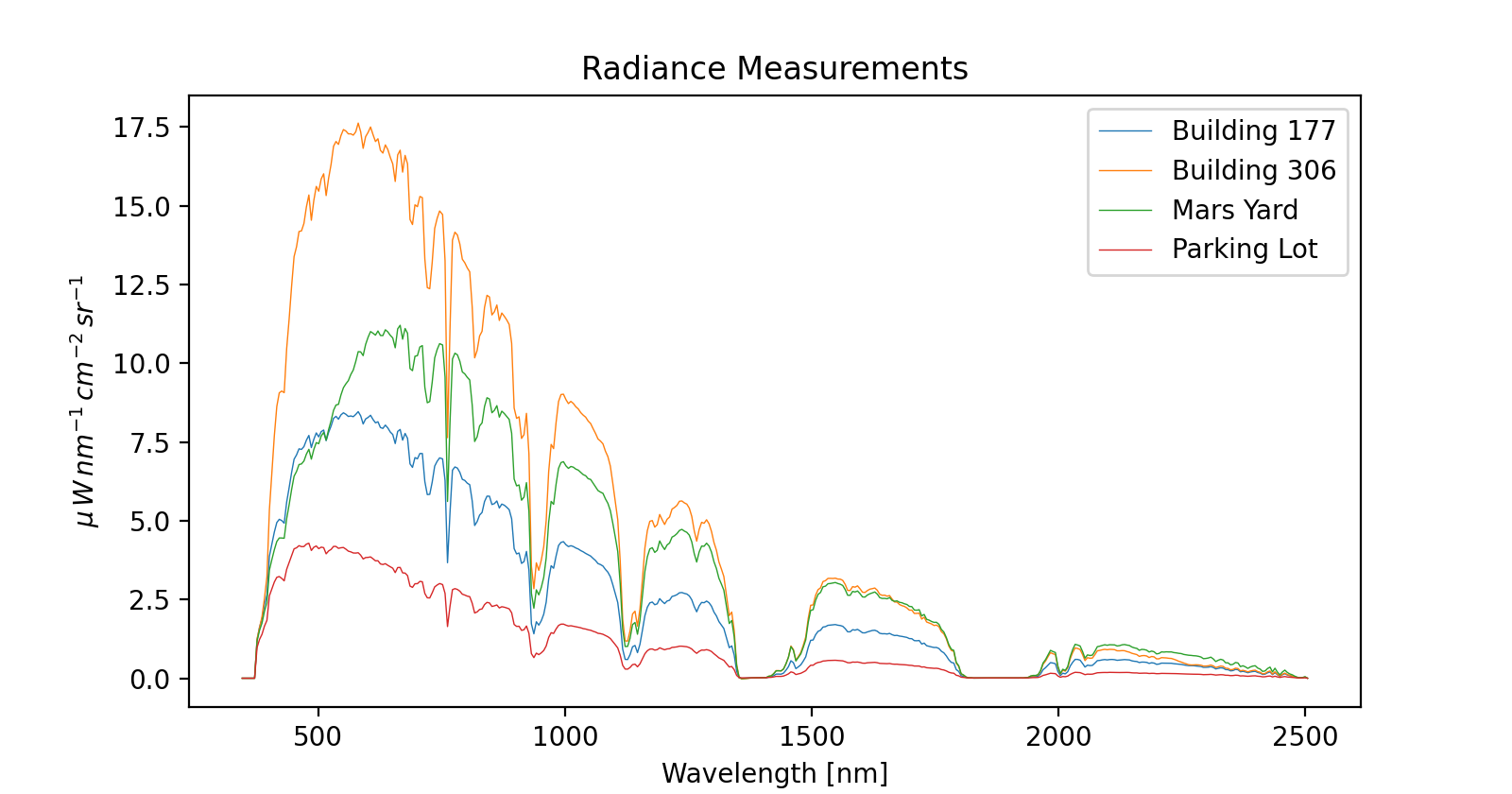}
    \caption{Radiance measurements for  test cases collected over JPL campus.}
    \label{fig:radiance}
\end{figure}

The results focus on comparing the posterior distribution characterized by the fully Bayesian MCMC method with the posterior approximated by optimal estimation. 
The MCMC algorithm was executed for four radiance datasets. 
The datasets, collected over the JPL campus using the airborne AVIRIS-NG instrument \cite{aviris1, aviris2}, attempt to include a variety of terrain types and are named Building 177, Building 306, Mars Yard, and Parking Lot. 
The radiance spectra are shown in Figure~\ref{fig:radiance}.
Two million samples were obtained for each chain, with the first $2\times 10^{5}$ discarded as burn-in since the chain stabilizes by then for most cases, as seen in the trace plots in Figure~\ref{fig:trace}.
The chain was thinned by taking every tenth sample to reduce storage.
The overall acceptance rate for all cases ranged from 0.2 to 0.3.

In this section, we compare the surface posterior obtained from both methods using several metrics, followed by the posterior on the atmospheric parameters.
Then, we evaluate the Gaussianity of the full posterior distribution.

\subsection{Surface posterior comparison}

\begin{figure}
    \centering
    \includegraphics[width=0.5\linewidth]{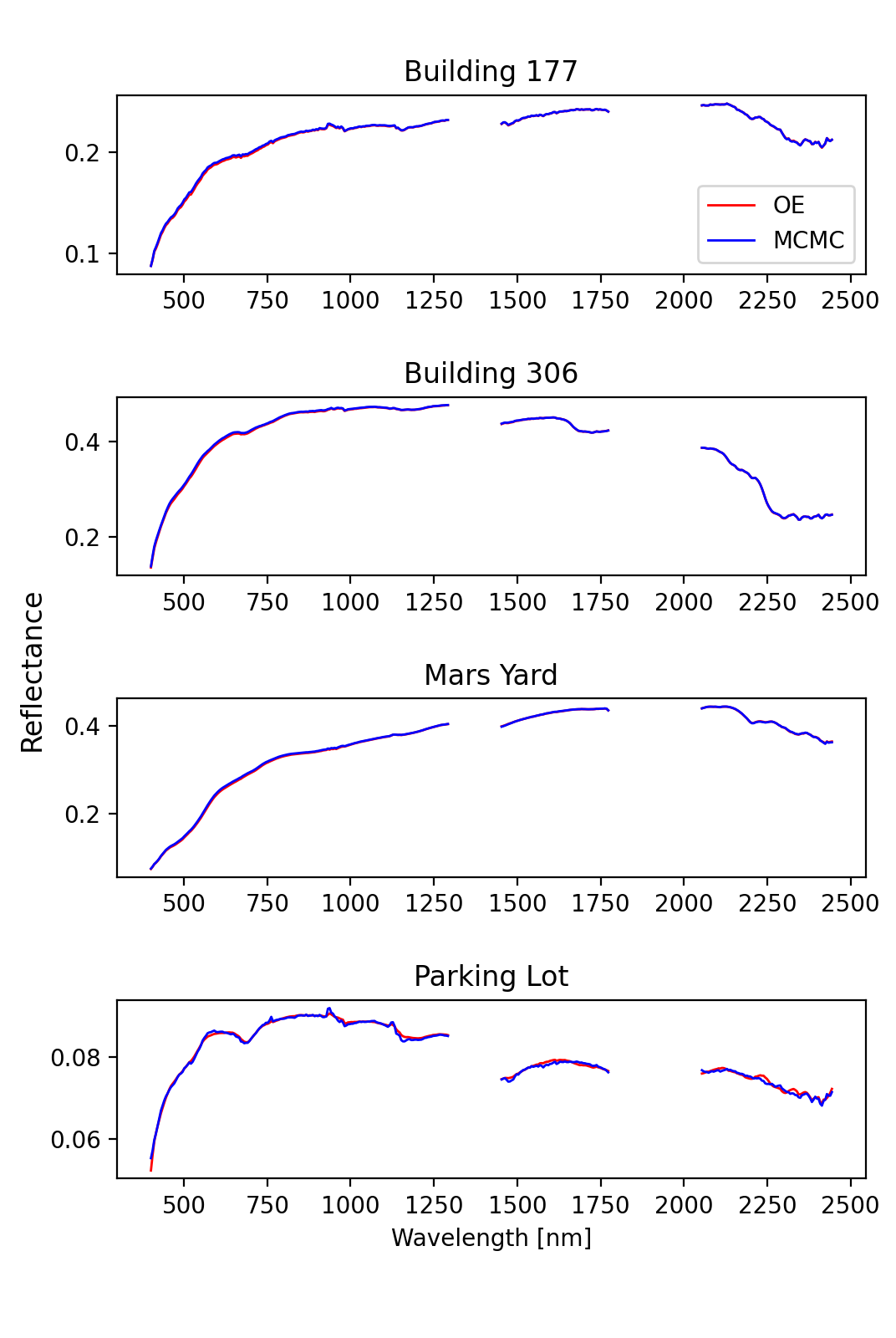}
    \caption{Posterior mean and MAP estimates for reflectances.}
    \label{fig:posmean}
\end{figure}

\begin{figure}
    \centering
    \includegraphics[width=0.5\linewidth]{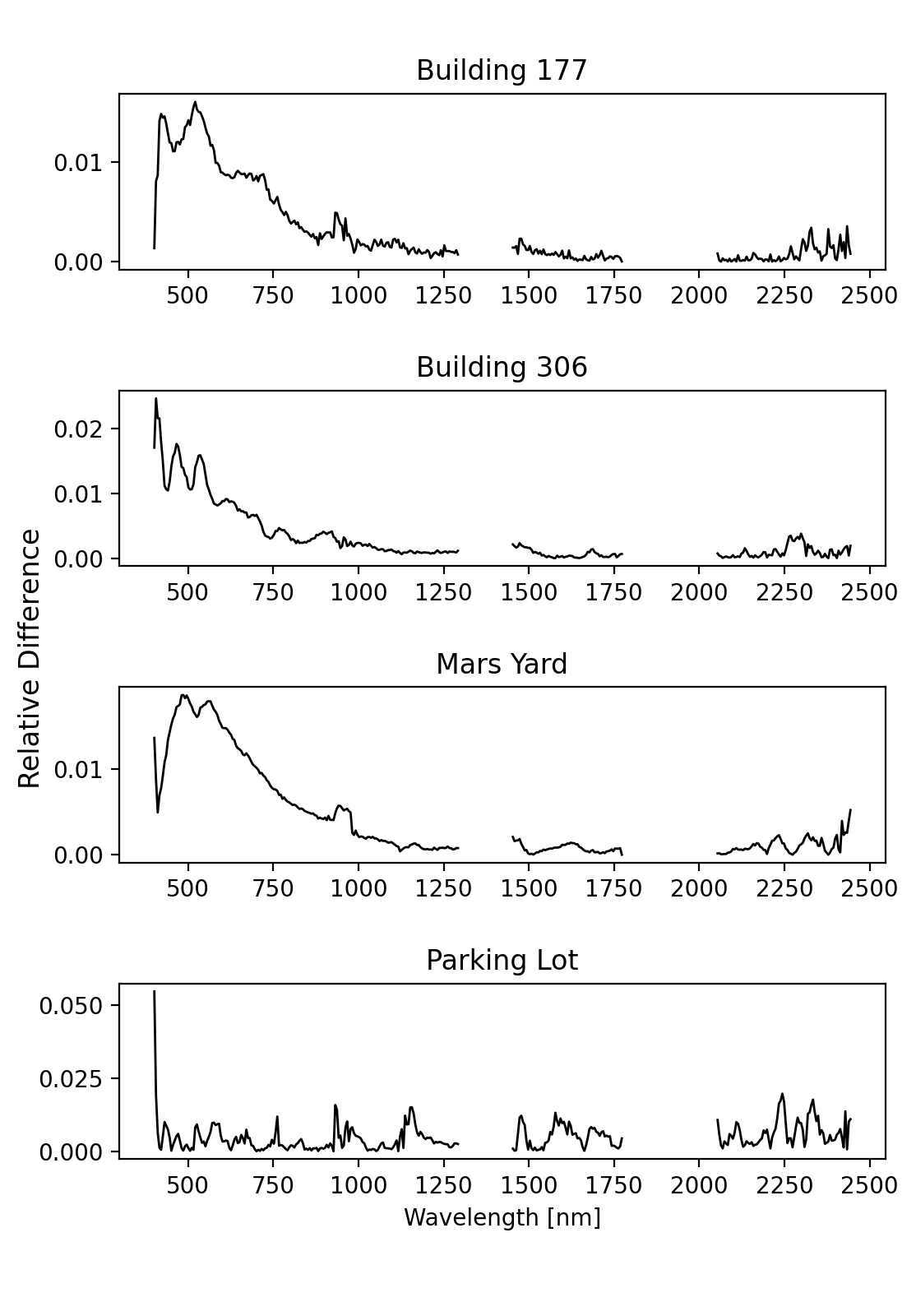}
    \caption{Relative difference between posterior mean and MAP estimates.} 
    \label{fig:posmeanerror}
\end{figure}

We first compare the mean and covariance of the posterior distribution. 
Figure~\ref{fig:posmean} plots the posterior mean reflectance for the four test cases and the MAP estimate from optimal estimation. 
Figure~\ref{fig:posmeanerror} plots the relative difference of the two methods normalized over the values from the MCMC method.
In the first three cases, the greatest differences occur in the low wavelength regions and all peak around 0.02. 
The Parking Lot case begins with a high relative difference at low wavelength but generally remain below 0.02.
The relative difference between the posterior mean and MAP estimates are below 6\% throughout the spectrum, which gives confidence in both methods that they converge to a sensible location in the parameter space. 

\begin{figure}
    \centering
    \includegraphics[width=0.5\linewidth]{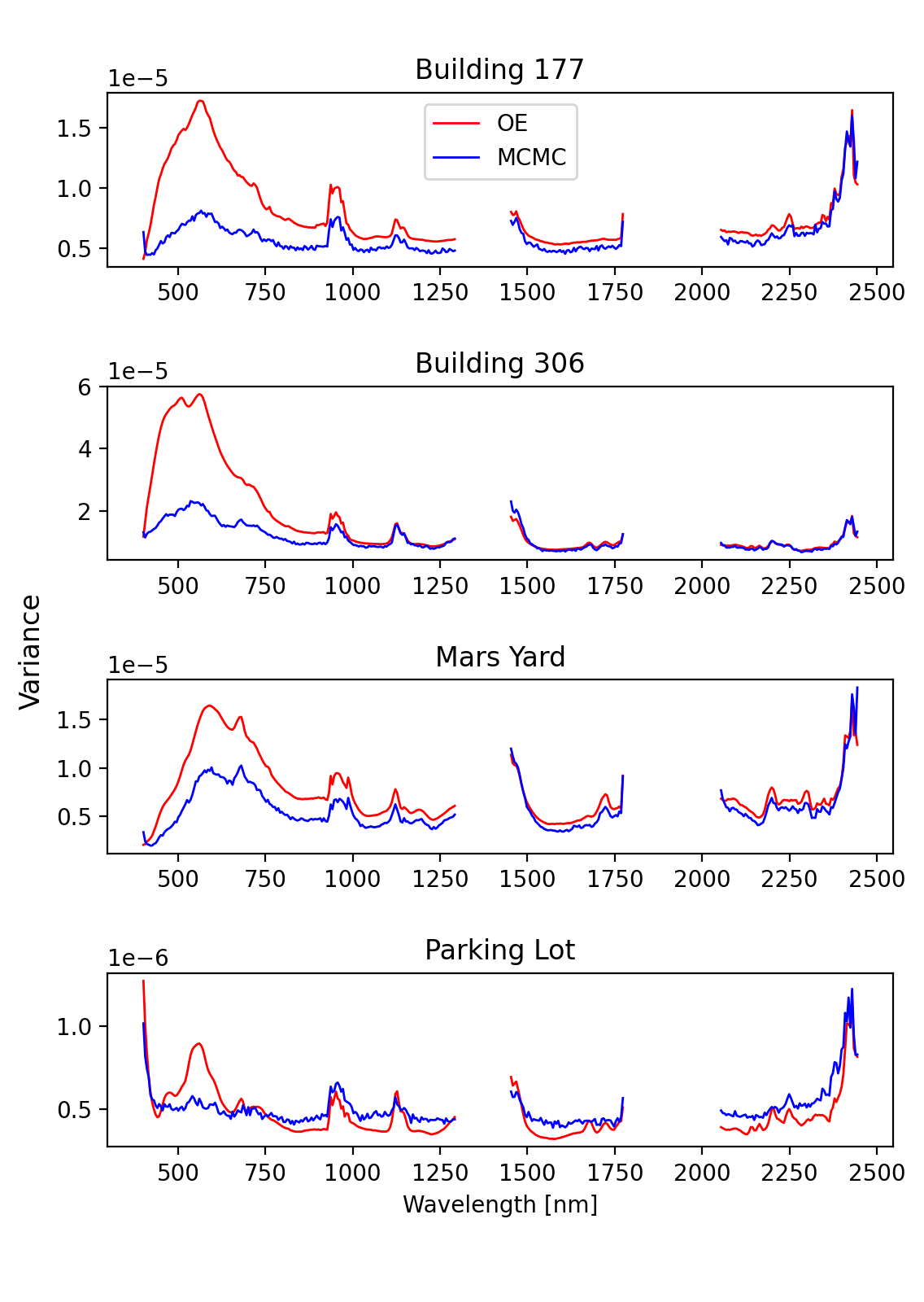}
    \caption{Marginal variance in reflectances predicted by MCMC and OE methods.}
    \label{fig:posvar}
\end{figure}

Figure~\ref{fig:posvar} shows the marginal variance of the posterior distribution in the reflectances obtained using Laplace approximation in optimal estimation and MCMC. 
For the first three cases, the Laplace approximation predicts a higher marginal variance than the MCMC posterior variance, particularly in the low wavelength regions below 1000 nm.
For the Parking Lot case, the posterior marginal variance is slightly higher than the Laplace approximation except for the regions around 380 nm and 550 nm. 

In addition to the marginal variances, it is also necessary to look into the differences in cross-correlations.
We compare the MCMC and OE covariances ($\Gamma_\text{M}$ and $\Gamma_\text{L}$) using three metrics involving trace, Frobenius norm, and Förstner distance, which are defined below.
The first metric is the relative difference in trace normalized by the trace of the MCMC posterior covariance:
\begin{equation}
    d_\text{tr} = \bigg| \frac{\text{tr} (\Gamma_\text{M}) - \text{tr} (\Gamma_\text{L}) }{\text{tr} (\Gamma_\text{M}) }\bigg|. 
\end{equation}
The second metric is the relative difference in Frobenius norm normalized by the Frobenius norm of the MCMC posterior covariance:
\begin{equation}
    d_\text{norm} = \frac{\|{\Gamma_\text{M} - \Gamma_\text{L} \|_F}}{\| \Gamma_\text{M} \|_F }.
\end{equation}
The third metric involves the Förstner distance, which is a metric that measures the distance between two symmetric positive definite matrices \cite{forstner}. 
The Förstner distance between two SPD matrices $\Gamma_A$ and $\Gamma_B$ is defined by:
\begin{equation}
    d_f =\sqrt{\sum \ln^2(\sigma_i)},
\end{equation}
where $\sigma_i$ are the generalized eigenvalues of the eigenpencil $(\Gamma_A, \Gamma_B)$.
The relative difference defined using this metric is normalized by the distance between the MCMC covariance and the prior covariance:
\begin{equation}
    d_F = \frac{ d_f\big(\Gamma_\text{M}\,,\Gamma_\text{L}\big) }{ d_f\big(\Gamma_\text{M}\,,\Gamma_\text{pr}\big)}.
\end{equation}

The results of these comparisons are shown in Table~\ref{tab:covdiff} for the four test cases.
The importance of using multiple metrics is highlighted in the Parking lot case, where the trace and Frobenius norm indicate a lower difference compared to the other three cases, but Förstner distance is higher than the other cases.
The relative difference between MCMC and Laplace approximation covariances are greater than 0.3 in all but three of the 12 values. 
This numerical comparison establishes that there is a significant deviation in the covariances obtained using the approximate Bayesian and fully Bayesian algorithms. 


\begin{table}
    \centering
    \caption{Relative difference between $\Gamma_\text{M}$ and $\Gamma_\text{L}$}
    \label{tab:covdiff}
    \begin{tabular}{|c|c|c|c|}
        \hline
         & Trace & Frobenius Norm & Förstner Distance \\
         \hline
        Building 177 & 0.331 &  1.820 & 0.319 \\
         \hline
        Building 306 & 0.509 & 2.601 & 0.282 \\
         \hline
        Mars Yard & 0.289 & 0.758 & 0.356 \\
         \hline
        Parking Lot & 0.063 & 0.929 & 0.539 \\
         \hline
    \end{tabular}
\end{table}


\subsection{Eigenanalysis of the surface posterior}

\begin{figure}
    \centering
    \includegraphics[width=0.5\linewidth]{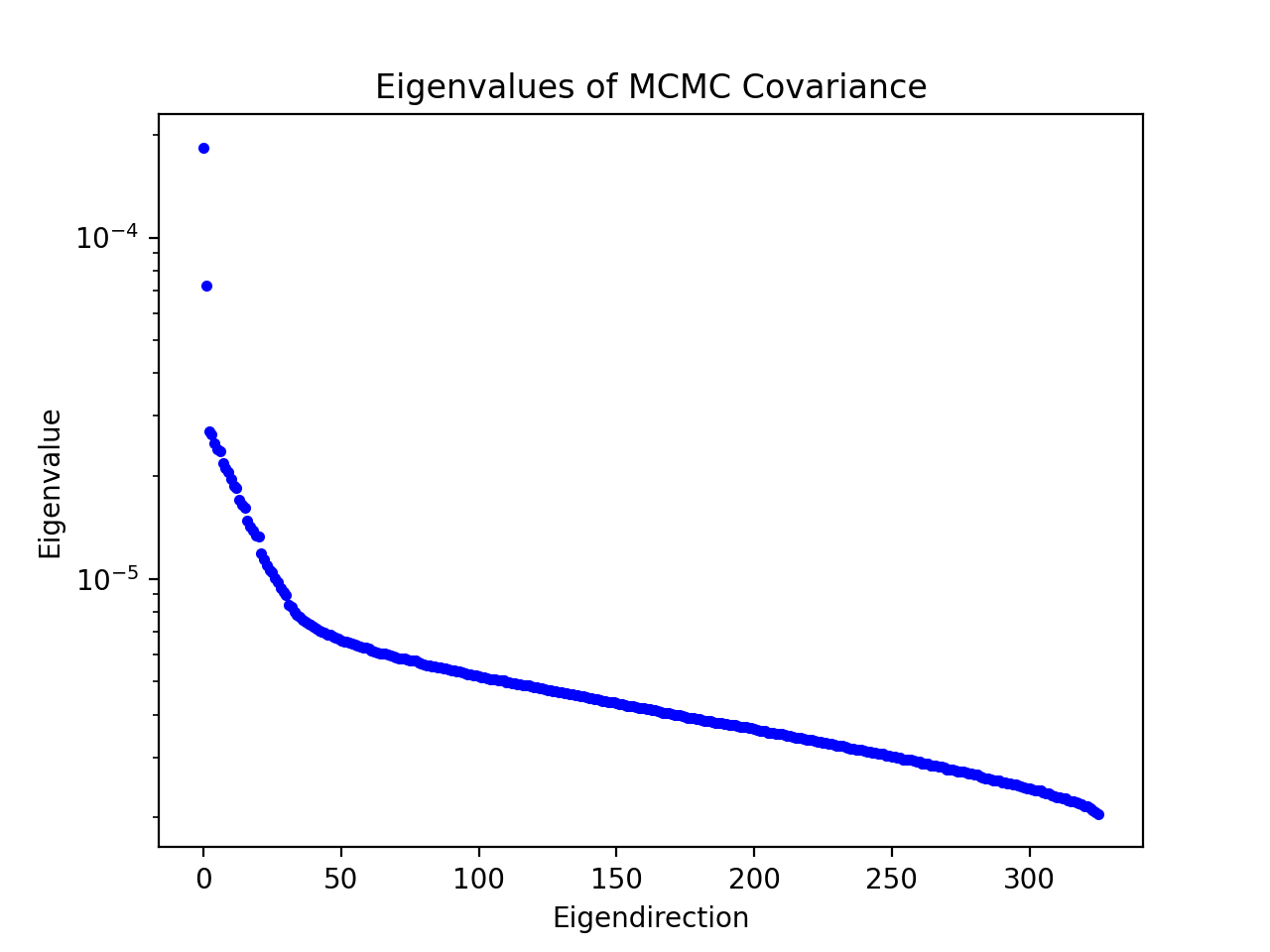}
    \caption{Eigenvalues of the MCMC covariance matrix.}
    \label{fig:eigval}
\end{figure}

Expanding on the eigenvalue problem used in the Förstner distance metric, we explore the interpretability of eigenproblems to reveal structure in the difference between the two covariance matrices.
We compare them with respect to the eigendirections of one of the matrices.
Specifically, we focus on the following eigenvalue problem involving the sample covariance of the Building 177 posterior:
\begin{equation}
    \Gamma_\text{M} v_\text{M} = \lambda_\text{M} v_\text{M},
\end{equation}
where $\lambda_\text{M}$ and $v_\text{M}$ are the eigenvalues and eigenvectors of $\Gamma_\text{M}$.
The eigenvalue spectrum is shown in Figure~\ref{fig:eigval}.

Then, the variance of $\Gamma_\text{L}$ in the direction $v_\text{M,i}$ can be expressed as $v_\text{M,i}^\top \Gamma_\text{L} v_\text{M,i}$.
This directional variance can be normalized using the corresponding eigenvalue as follows:
\begin{align}
\begin{split}
    \sigma^\text{L}_\text{M,i} &= \frac{v_\text{M,i}^\top \Gamma_\text{L} v_\text{M,i}} {v_\text{M,i}^\top \Gamma_\text{M} v_\text{M,i}} = \frac{v_\text{M,i}^\top \Gamma_\text{L} v_\text{M,i}} {\lambda_\text{M,i}}
\end{split}
\end{align}

\begin{figure}
    \centering
    \includegraphics[width=0.7\linewidth]{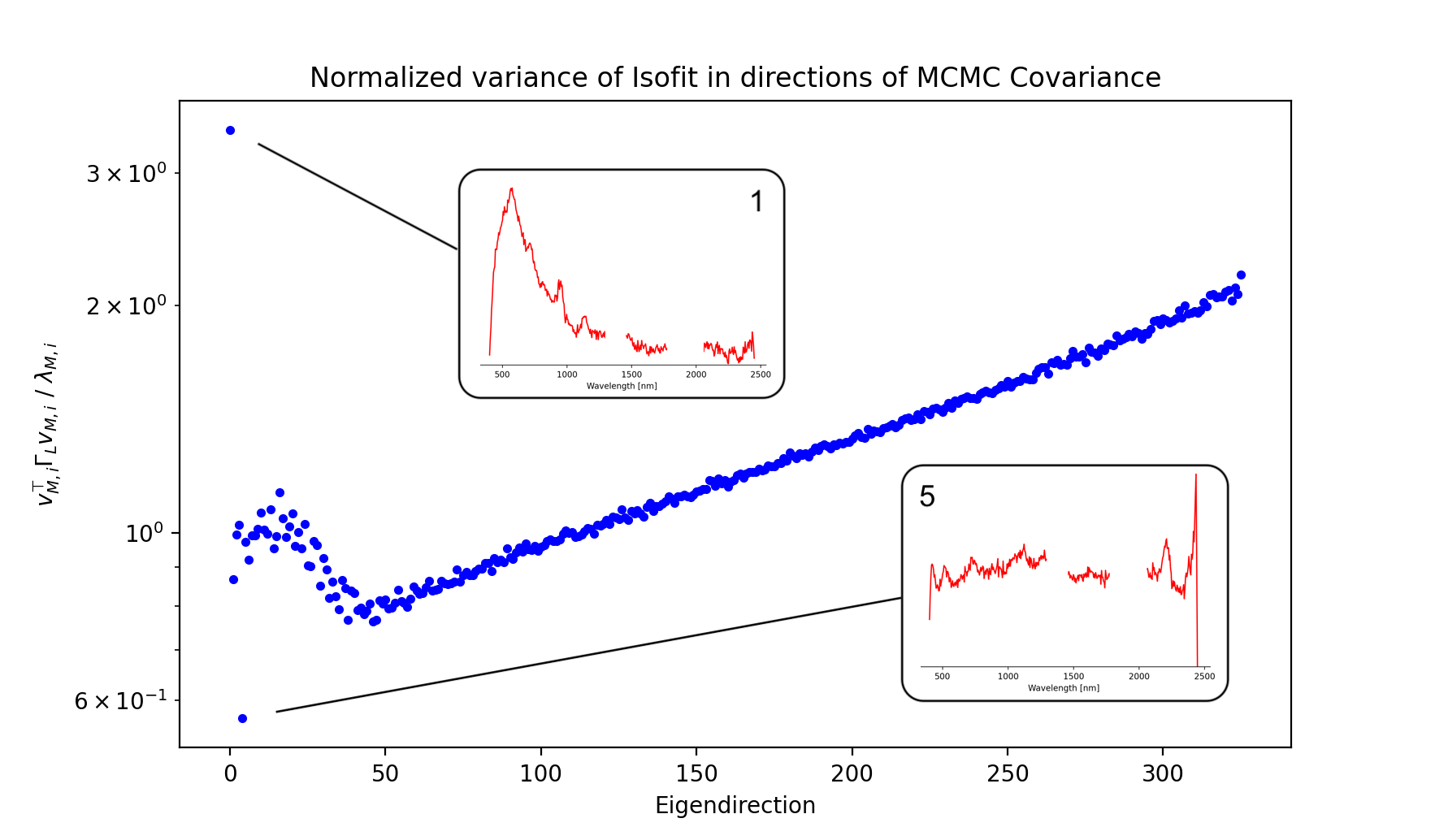}
    \caption{Ratio of Laplace approximation variance and MCMC variance in the eigendirections of the MCMC covariance.}
    \label{fig:quotient}
\end{figure}

Figure~\ref{fig:quotient} plots this quotient, ranked in order of highest to lowest eigenvalue $\lambda_\text{M,i}$.
A value greater than 1 can be interpreted as the Laplace approximation having greater variance in the $v_\text{M,i}$ direction, and vice versa.
Consistent with Figure~\ref{fig:posvar}, the Laplace approximation overestimates the variance of the MCMC variance in most directions.
The overall pattern is that the variances are around the same for most of the more important eigendirections, but the MCMC variance becomes smaller than the Laplace approximation variance going into the less important directions.


Two interesting points to further analyze are the outliers near the leading eigendirections in Figure~\ref{fig:quotient}.
The corresponding eigenvectors are shown in red.
Comparing the shape to the posterior variance plot in Figure~\ref{fig:posvar}, the first eigenvector (top outlier) resembles the main feature in the lower wavelengths.
The Laplace approximation predicts variance three times higher in this direction.
In Section~\ref{subsec:gaussianity} we show how this may be related to the non-Gaussianity in the low wavelength region, which makes the Laplace approximation less accurate.
The fifth eigenvector (bottom outlier) describes some of the noisy features, particularly the spike near 2500 nm, and the MCMC result predicts a variance around 70\% higher than the Laplace approximation in this direction.



\subsection{Atmospheric posterior comparison}

While the reflectances are the quantities of interest ultimately used in subsequent analysis of the Earth surface, their behaviour is conditioned on the atmospheric parameters.

\begin{figure}
    \centering
    \includegraphics[width=0.6\linewidth]{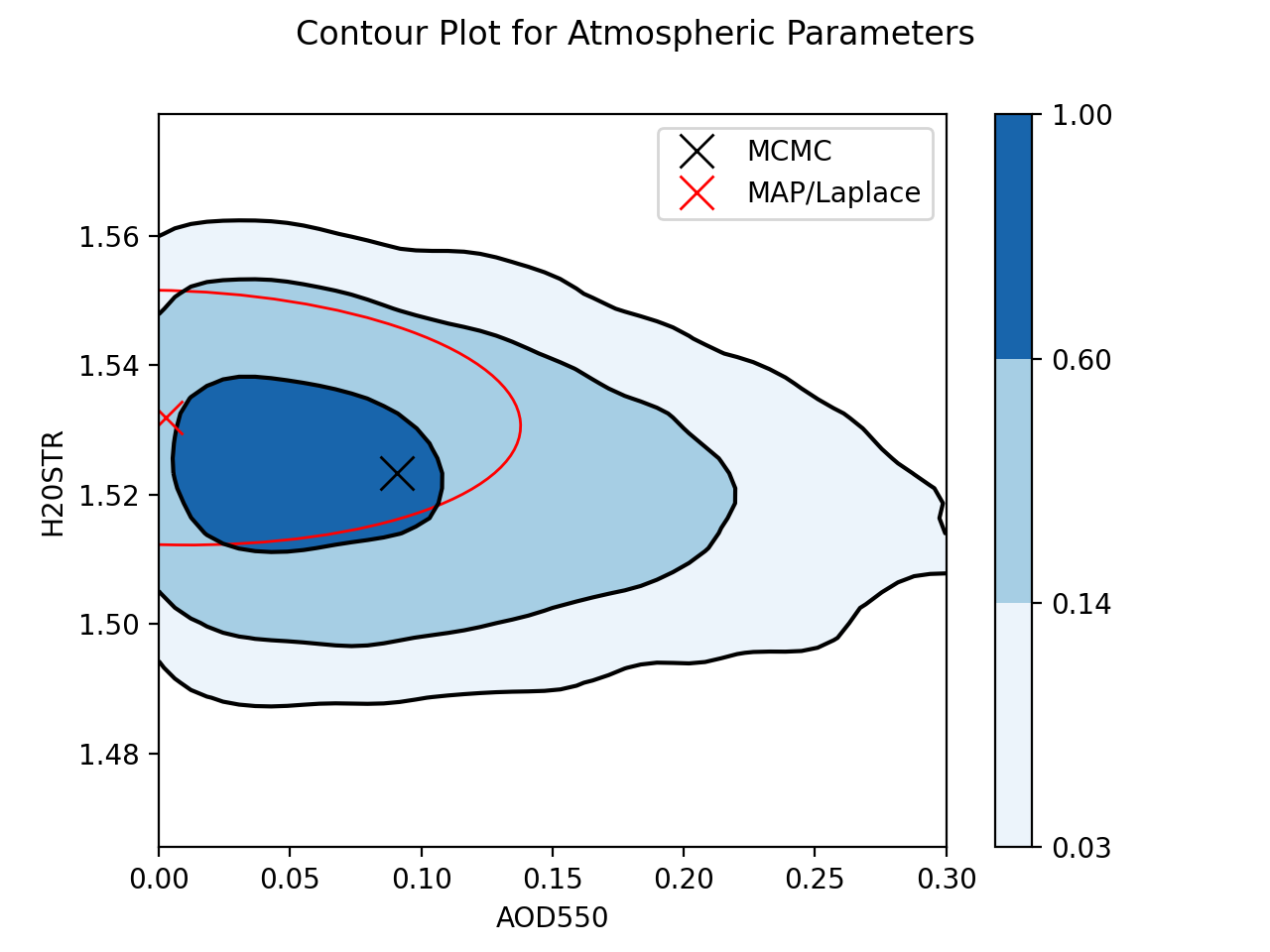}
    \caption{2D marginal density plot of the atmospheric posterior distribution for Building 177.} 
    \label{fig:contouratm}
\end{figure}

Figure~\ref{fig:contouratm} is a 2D marginal density plot of the posterior for the two atmospheric parameters.
The MAP estimate from optimal estimation is plotted in red along with an ellipse representing one standard deviation obtained using the Laplace approximation.
There are two visible improvements from characterizing the posterior distribution using MCMC.
First, optimal estimation has no way of ensuring positivity of the parameters, so the probabilistic interpretation is that the probability is obtaining a negative AOD parameter is almost 0.5, for example.
The MCMC implementation fixes the samples to be positive and therefore leads to results that are more representative of the physical quantities. 
The second improvement is that MCMC sampling reveals a non-elliptical shape to the posterior, suggesting that it is not Gaussian. 
The Gaussianity of the posterior for both surface and atmospheric parameters is further explored in the next subsection.

\subsection{Evaluating Gaussianity} \label{subsec:gaussianity}

The motivation for turning to a fully Bayesian approach is that the posterior is non-Gaussian in general.
Here, we first demonstrate the non-Gaussianity of the posterior distribution qualitatively using normal Q-Q plots, and then quantitatively using hypothesis testing for individual parameters in one dimension.

\begin{figure}
    \centering
    \includegraphics[width=0.5\linewidth]{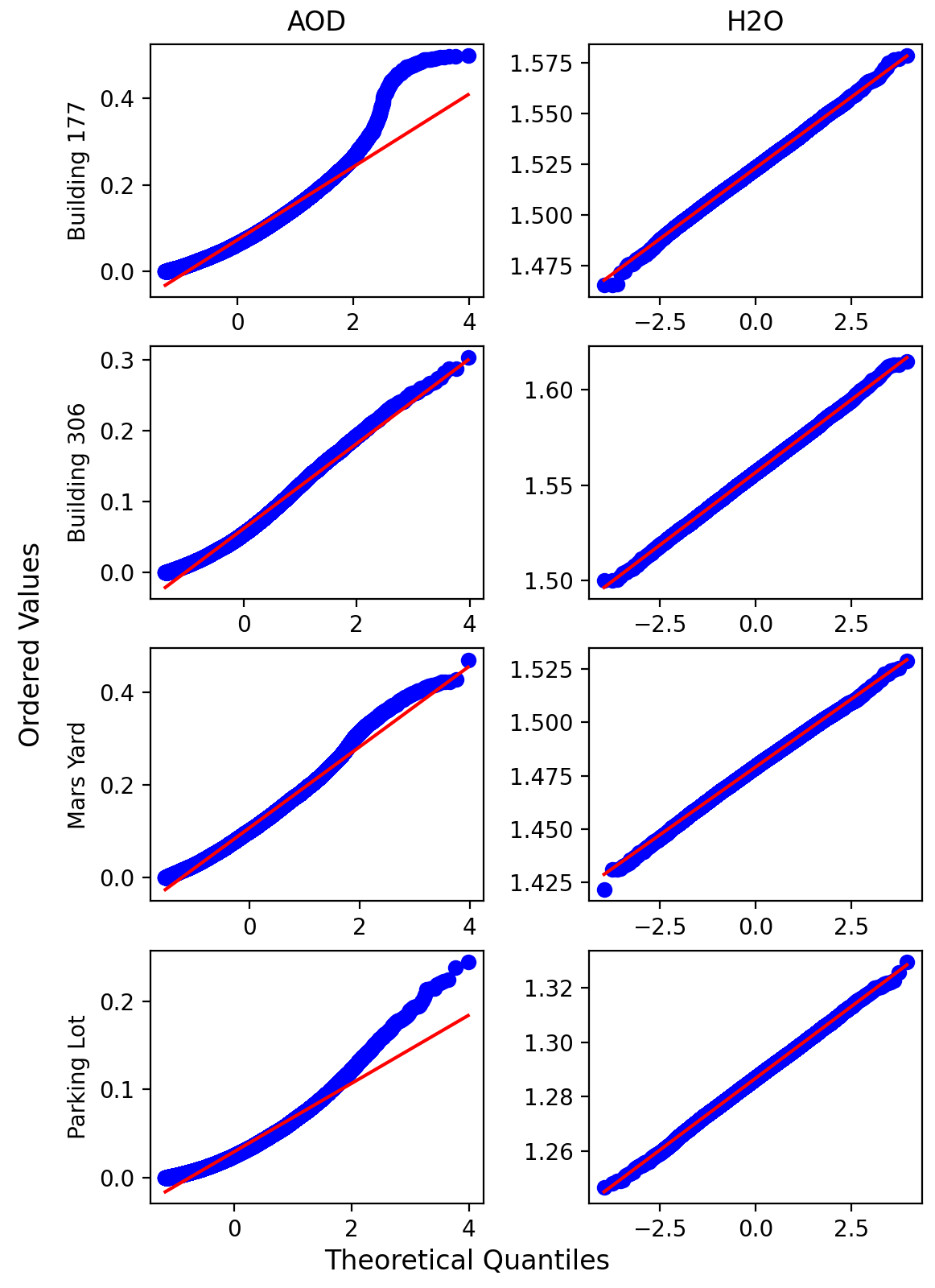}
    \caption{Q-Q plots of atmospheric parameters across all four cases. The red line is a reference indicating the truncated normal distribution.}
    \label{fig:qqatm}
\end{figure}

Figure~\ref{fig:qqatm} shows the Q-Q plots for the two atmospheric parameters across all four cases.
The red line is the reference for a truncated normal distribution and the MCMC samples are plotted in blue.
The truncated normal was used here since the samples for the atmospheric block were constrained to positive values in the algorithm. 
While the \ch{H2O} parameters closely follow the truncated normal, the right tail of the AOD plots deviate from the red line, especially for the Building 177 and Parking Lot cases.
Qualitatively, these two cases look the least Gaussian. 

\begin{figure}
    \centering
    \includegraphics[width=0.5\linewidth]{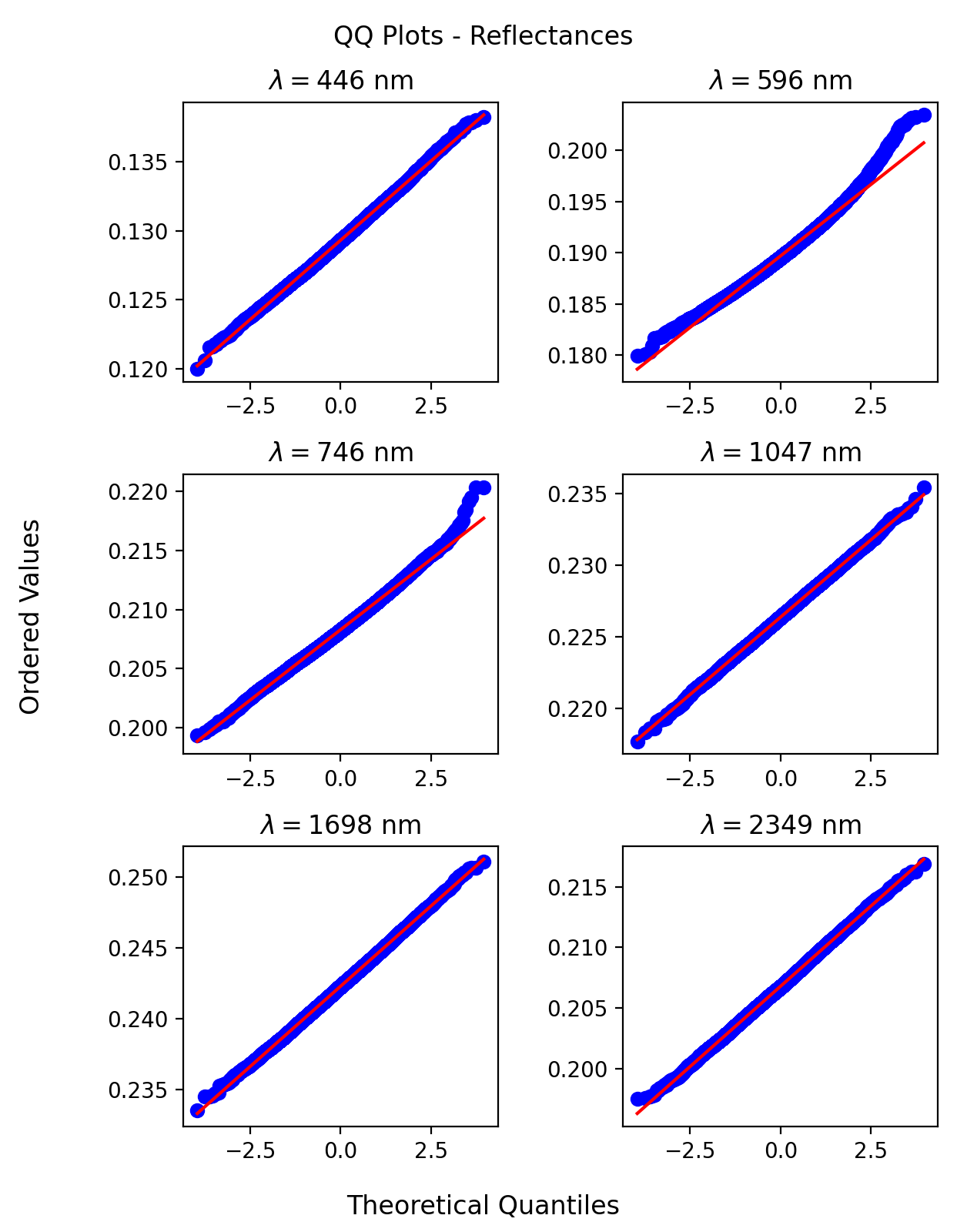}
    \caption{Q-Q plots of select reflectance parameters for the Building 177 case. } 
    \label{fig:qqref}
\end{figure}

Figure~\ref{fig:qqref} shows the Q-Q plots for select reflectance parameters across the spectrum for the Building 177 case.
Although the MCMC samples closely follow a normal distribution, the two plots for 596 nm and 746 nm have tails that deviate from the reference normal.

\begin{figure}
    \centering
    \includegraphics[width=0.7\linewidth]{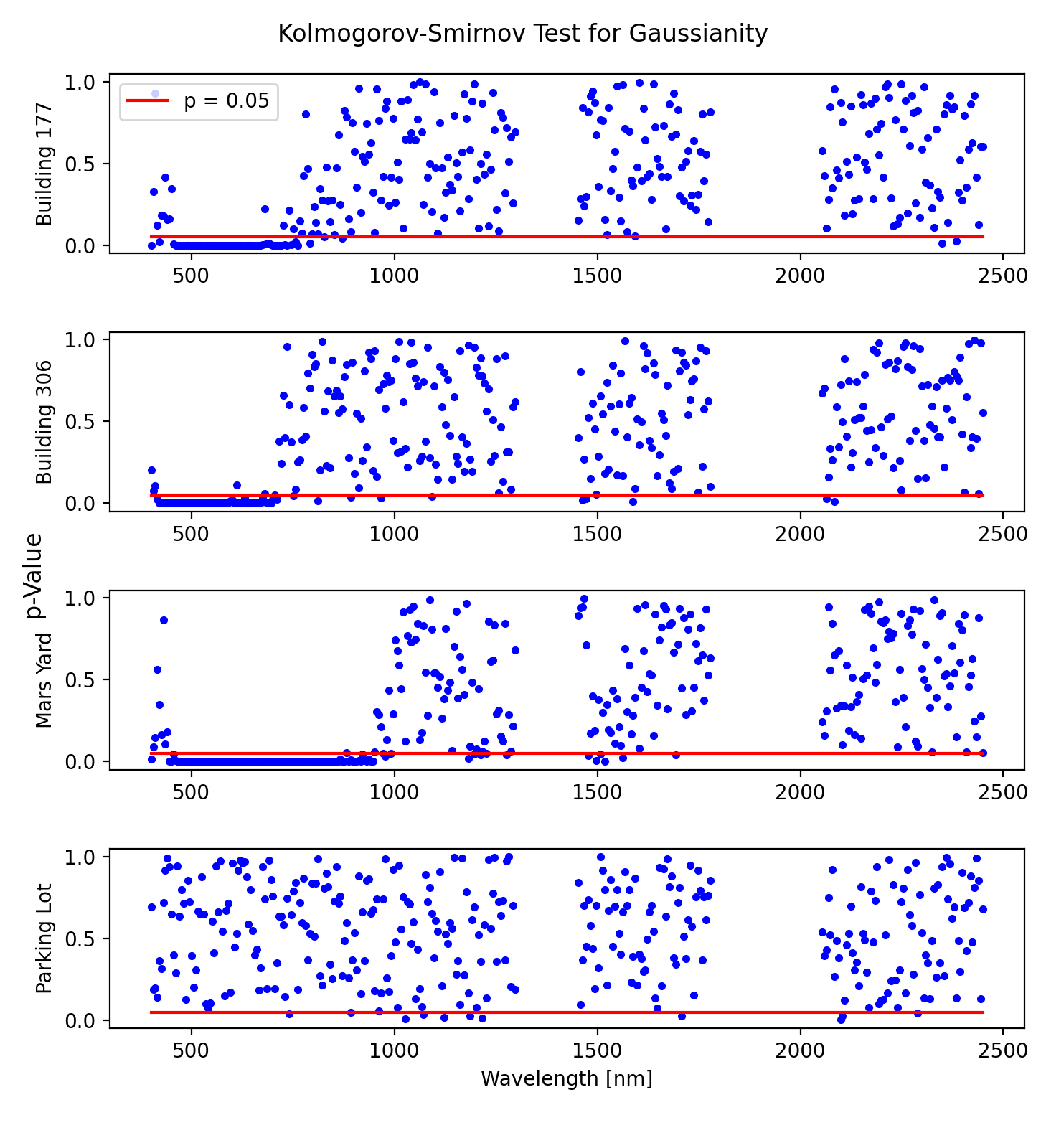}
    \caption{Kolmogorov-Smirnov test for reflectances across all four cases.} 
    \label{fig:ksref}
\end{figure}

Next, we present a more comprehensive analysis of the reflectances using a hypothesis testing approach.
Treating the reflectances individually, we use the Kolmogorov-Smirnov test \cite{kstest} on the empirical marginal distribution of the MCMC samples with the null hypothesis being that the reflectances are normally distributed.
The $p$-values for each reflectance parameter are shown in Figure~\ref{fig:ksref}, with the red line representing $p=0.05$.
In three of the cases, the non-Gaussian phenomenon observed in certain wavelengths in Figure~\ref{fig:qqref} are present in the entire low wavelength regime, with $p\approx 0$.
However, the extent of this regime varies for all three cases, with the low $p$-value region in the Mars Yard case extending to nearly 1000 nm.

This is consistent with the findings in Figure~\ref{fig:quotient}, which showed that the largest difference between the OE and MCMC posterior was in the direction that represents the lower wavelength regime. 
The departure from Gaussianity for the reflectances in this regime may be the reason why the Laplace approximation also departs from the posterior characterized by MCMC.



\section{Discussion}

We presented a fully Bayesian MCMC algorithm for the remote sensing problem that characterizes the posterior distribution of the surface reflectances and atmospheric parameters.
This posterior was used to identify and understand the limitations of optimal estimation, the current state-of-the-art approximate Bayesian approach.
There are three main takeaways from the results presented in this paper.
\begin{itemize}
    \item The fully Bayesian solution and approximate Bayesian solution yield very different covariances. We analyzed the differences in terms of three different metrics and the eigendirections of the MCMC covariance matrix.
    \item The posterior distribution of the atmospheric parameters is more physically sensible than the Laplace approximation.
    \item Non-Gaussianity in the posterior is revealed by the fully Bayesian solution.
\end{itemize}
We identified regions of the spectrum and in the atmospheric parameters for which the Laplace approximation would not be able to sufficiently represent the non-Gaussian distribution.
From the eigenanalysis, the OE posterior covariance was shown to be the most different from the MCMC posterior covariance in the low-wavelength region, which is the same region that was shown to depart from Gaussianity.
Any further work on non-Gaussian posterior characterizations could focus on only this regime of the reflectances and the AOD atmospheric parameter. 
There is potential to develop a new combined method that uses MCMC or another non-Gaussian method for these parts, and OE for the rest of the parameters.


In terms of NASA's Surface Geology and Biology mission, characterizing the posterior distribution is important for subsequent analysis.
The surface reflectances are ultimately used to further infer properties of the Earth surface pertaining to problems such as ecosystems and ice accumulation.
Accurately quantifying the uncertainty of the surface properties is especially important for these scientific applications.


\subsection{Limitations}

Although we have created a computationally tractable fully Bayesian algorithm by exploiting structure in the problem, it is not computationally feasible in an operational setting.
Generating $2\times 10^6$ posterior samples takes on the order of hours, whereas the approximate Bayesian approach for one retrieval is on the order of seconds.
However, the two methods could be combined for the operational setting. 
For example, a fully Bayesian retrieval can be used as a validation step to verify and potentially correct the Laplace approximation.
Or, since the Laplace approximation is sufficient to approximate many of the parameters, the algorithm in this work can be narrowed to solve a smaller subproblem on a subset of parameters using a fully Bayesian approach while maintaining the original approximate approach for the other parameters.

The performance of the MCMC algorithm is contingent on the prior, noise model, and forward model.
Since the noise model dictates the likelihood distribution, the posterior would change along with changes in the prior distribution and noise model.
The forward model is subject to modelling error, which could affect the interactions between the atmospheric and surface parameters.
This would be one of the main improvements to the current algorithm when considering operational use, since ultimately we are interested in matching the ground truth reflectances.


\subsection{Future work}

The retrievals performed in this work are pixel-by-pixel, meaning that for each radiance vector, there is one corresponding state vector to be inferred. 
The main efforts for future work are focused on extending fully Bayesian algorithms to spatial and temporal fields. 
Including spatial and temporal correlations can increase the retrieval accuracy and reduce the number of retrievals required, which would reduce computational time in an operational setting.


\section{Conclusion}
In this work, we developed a computationally tractable fully Bayesian retrieval method for the high-dimensional VSWIR retrieval.
Taking advantage of the structure in the forward radiative transfer model, we implemented a block Metropolis MCMC algorithm that alternates samples between the atmospheric and surface parameter blocks.
Unlike other algorithms that use dimension reduction, the block Metropolis algorithm is asymptotically exact. 

We compared this fully Bayesian algorithm to the current state-of-the-art optimal estimation method in several ways to identify limitations of the approximate Bayesian approach.
For the surface parameters, both methods had a similar posterior mean but the posterior variance was shown to be significantly different.
The eigendirections in which they are different were analyzed and interpreted.
MCMC revealed non-Gaussianity in the aerosol parameter and the low-wavelength regime of the surface reflectances, determined through Kolmogorov-Smirnov hypothesis tests.
This work is the first step in combining the block Metropolis algorithm with optimal estimation to allow non-Gaussian retrievals to be performed in an operational setting.

\section*{Acknowledgments}
A portion of this research was performed at the Jet Propulsion Laboratory, California Institute of Technology, under contract with the National Aeronautics and Space Administration. This research was funded in part by a Jet Propulsion Laboratory Strategic University Partnership grant to MIT. Kelvin Leung and Youssef Marzouk also acknowledge support from the Office of Naval Research, SIMDA (Sea Ice Modeling and Data Assimilation) MURI, award number N00014-20-1-2595.

\bibliographystyle{unsrt} 
\bibliography{references}

\end{document}